\documentclass[showpacs,preprintnumbers,amsmath,amssymb]{revtex4}

\usepackage{graphicx}
\usepackage{dcolumn}
\usepackage{bm}

\begin{document}


\title{Stress orientation of second-phase in alloys: hydrides in zirconium alloys
\footnote{A portion of this paper was presented in MS\&T'08, October 5-9, 2008, Pittsburgh,
Pennsylvania, USA.}} 
\author{ A. R. Massih$^{a,b}$ and L. O. Jernkvist$^{a}$}%
\affiliation{%
$^{a}$Quantum Technologies, Uppsala Science Park, SE-75183 Uppsala, Sweden\\
$^{b}$Malm\"{o} University, SE-20506 Malm\"{o}, Sweden}

\date{\today}

\begin{abstract}
A model for precipitation of the plate-shaped second-phase under
applied stress is presented. The precipitates in the
matrix-precipitate system are represented by their local volume
fraction $\zeta$ and an orientation parameter $\theta$ that defines
the alignment of a precipitate platelet in a given direction. Kinetic
equations, based on diffusion theory and classical nucleation theory,
are used to describe the time evolution of $\zeta$ and $\theta$ . The
model is used to describe the stress orientation of hydrides in
Zr-alloys in light of experiments.
\end{abstract}

\maketitle

\section{Introduction}
\label{sec:intro}
Stress-induced orientation of second-phase particles occurring during
precipitation is ubiquitous in many metallurgical systems.  Stress
orientation is referred to the selective precipitation of a
second-phase on a preferred habit plane, caused by the application of
an external stress during aging or cooling. For example, in the Ti-H
system, hydride precipitation has been observed to occur under an applied
tensile stress, upon which titanium hydrides form on the titanium habit
plane oriented most nearly perpendicular to the stress axis
\cite{Louthan_1963}. Similarly, in polycrystalline zirconium alloys,
it has been observed that stress-orientation of hydrides occurs under
tensile stresses, where hydride platelets would get aligned
perpendicular to the stress axis and can lie in specific
habit-planes; namely on either $\{10\bar{1}l\}$ or $\{11\bar{2}l\}$
crystal planes \cite{Louthan_Angerman_1966}. Under compressive
stresses, however, hydride platelets tend to orient parallel to
the direction of applied stress \cite{Louthan_Marshall_1963}.

The stress orientation effect has also been observed in Fe-N alloys,
in which Fe$_{16}$N$_2$ platelets ($\alpha^{\prime\prime}$ phase)
precipitate on a specific habit plane perpendicular to the direction
of applied tensile stress \cite{Nakada_et_al_1967,Tanaka_et_al_1978}.
Other alloy systems exhibiting stress orientation effects include
Ni-base alloys with Ni-Nb precipitates \cite{Oblak_et_al_1974},
Fe-Mo-Au with Au particles \cite{Sauthoff_1977}, and Al-Cu alloys with
$\theta^\prime$ second-phase particles
\cite{Hosford_Agrawal_1975,Eto_et_al_1978,Skrotzki_et_al_1996}.
Similarly, favoured orientation of precipitates has been observed in
ferromagnetic materials, when the material is cooled or heat-treated
below its Curie temperature in magnetic field. For instance, magnetic
field orienting of Fe$_{16}$N$_2$ platelets in Fe-N solid solution and
alignment of disc-shaped nitrogen atom clusters on the \{001\} matrix
planes of a nitrogen ferrite alloy in the presence of magnetic field
\cite{Neuhauser_Pitsch_1971,Ferguson_Jack_1985,Sauthoff_Pitsch_1987}.

 The precipitation of titanium hydride in titanium and zirconium
hydride in zirconium alloys indicates that the stress orientation of
the hydride platelets occurs through orientation of nuclei rather than
through selective growth after the critical nucleus size has been
reached
\cite{Louthan_1963,Louthan_Angerman_1966,Louthan_Marshall_1963,Ells_1970,Hardie_Shanahan_1975,Sakamoto_Nakatsuka_2006}.
Likewise, quantitative metallographic studies indicate that
$\theta^\prime$-phase alignment in Al-Cu alloys occurs during
nucleation and not under the subsequent growth
\cite{Eto_et_al_1978,Skrotzki_et_al_1996}.

In this paper, we present a model for kinetics of precipitation of
second phase in solid solution under an applied stress. The model may
be considered as a mean field description of the matrix-precipitate
system, in which the plate-like precipitates are characterized by
their volume fraction and a parameter describing their orientation in
the matrix. We further assume that second-phase orientation occurs
during the nucleation stage of the precipitation process. Section
\ref{sec:model} describes the model for precipitation kinetics and
stress orientation. In section \ref{sec:zr-alloy}, a case of hydride
precipitation and orientation in a zirconium alloy is considered,
where the material-dependent parameters entering the model are
quantified in light of experimental data. The paper concludes with a
discussion of results.

\section{Model}
\label{sec:model}

\subsection{Precipitation kinetics}
\label{sec:precip}
A simple geometrical model for the plate-shaped precipitates in solid
solution is envisioned. The precipitates consist of a regular array of
parallel ellipsoids embedded in a metal matrix. The sample material,
initially at some high temperature, is cooled below the solubility
line. Precipitation proceeds by diffusion of solute atoms from the
metal matrix to locations, e.g. dislocation lines and/or grain
boundaries, where second-phase particles form. A model analogous to that used by
Ham \cite{Ham_1959} for describing precipitation of solute atoms on
dislocations is considered (see appendix \ref{app:A}). In this model,
the precipitated fraction of excess solute $w$ can be calculated as a
function of time $t$ fairly accurately by the simple relationship of
the form: $w \cong 1-e^{-t/\tau_0}$,
where $\tau_0=\ell^2/(\alpha_0^2D)$, $\ell$ is the inter-precipitate
distance, $\alpha_0$ the lowest
eigenvalue of the prevailing diffusion equation, and $D$ is the solute
diffusivity. Here $w$ is a solution of
the kinetic equation
\begin{equation}
  \label{eq:Hydride_fraction_eq}
  \frac{dw}{dt} + \frac{1}{\tau_0}(w-1) = 0.
\end{equation}
\noindent
It is worth noting that Eq. (\ref{eq:Hydride_fraction_eq}) pertains to
a class of kinetic equations used to describe the overall
precipitation process of excess solute  \cite{Jain_&_Hughes_1978}.
Now in our convention, we put $w=\zeta/\zeta_e$, where $\zeta_e$ is
the precipitate volume fraction in equilibrium, i.e. after infinite
time if the current temperature and stress state were held constant. Hence
Eq. (\ref{eq:Hydride_fraction_eq}) is transformed to
\begin{equation}
  \label{eq:Hydride_fraction_app}
  \frac{d\zeta}{dt} + \frac{1}{\tau_0}\left( \zeta-\zeta_{e} \right) = 0.
\end{equation}
\noindent
which gives the rate of change for the precipitate volume fraction at a
given point in the material. Equation (\ref{eq:Hydride_fraction_app})
can be solved, subject to an initial condition, $\zeta(t=0)=\zeta_0$,
and the condition that $\zeta_e$ and $\tau_0$ are constant, yielding
\begin{equation}
  \label{eq:Hydride_fraction_sol}
  \zeta(t) = \zeta_{e}\Big[1-\Big(1-\frac{\zeta_0}{\zeta_e}\Big)e^{-t/\tau_0}\Big].
\end{equation}
\noindent
We consider now a material in which the
solubility limit for precipitation is identical to the solubility for
dissolution. The solute solubility is thus assumed to be independent of the
heating/cooling history and depends only on the current state of the
material. We should note that  $\zeta_{e}$  is a unique function of
concentration, temperature $T$ and pressure $p$ in the
considered material, and can be calculated from the phase diagram
by use of the lever rule
\begin{equation}
  \label{eq:Lever_rule}
  \zeta_{e} = \frac{C-C_{L}}{C_{U}-C_{L}},
\end{equation}
where $C$ is the total solute concentration in the material and
$C_{L}$ and $C_{U}$ denote the respective lower and upper boundary of
the mixed phase region of the phase diagram (temperature
vs. concentration).  Hence, $C_{L}$ is the solubility limit in the
matrix phase. Equation (\ref{eq:Lever_rule}) is valid for $C_{L} \le C
\le C_{U}$. It follows that our assumption of constant $\zeta_e$ in
deriving Eq. (\ref{eq:Hydride_fraction_sol}) implies that temperature,
pressure and solute total concentration are supposed to be
non-varying.

\subsection{Stress orientation}
\label{sec:orient}
\begin{figure}[htbp]
\begin{center}
    \includegraphics[width=0.75\textwidth]{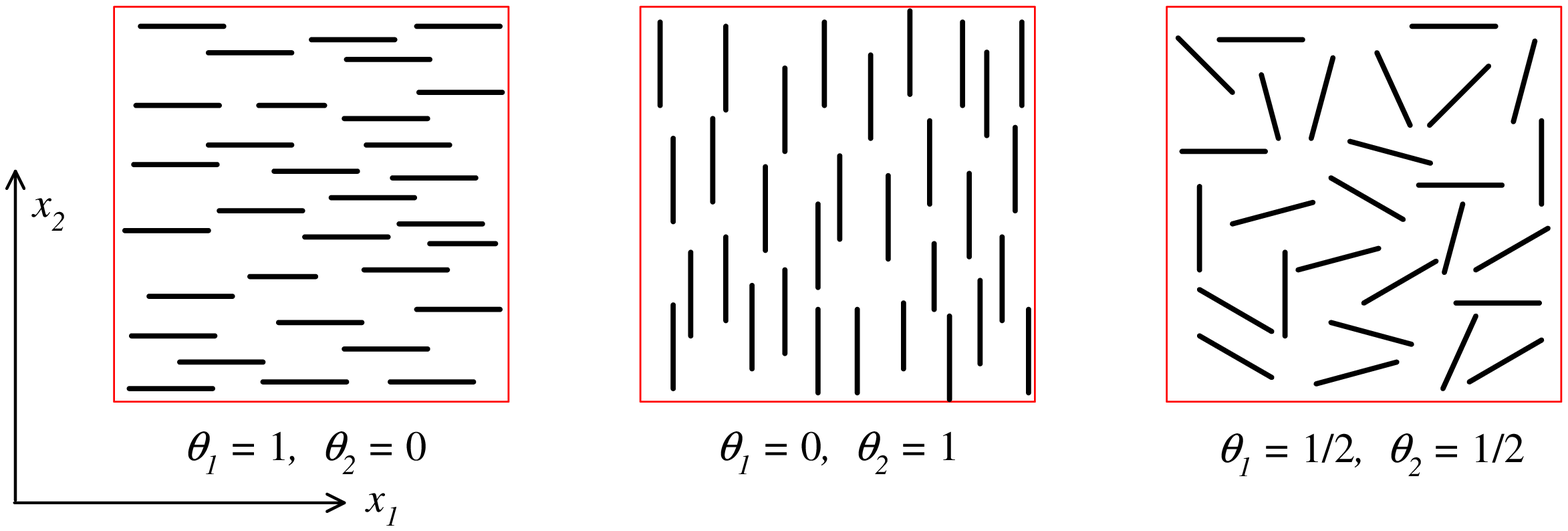}{} \\
  \end{center}
    \caption{Two-dimensional illustration of the variables
      $\theta_{i}$, which define the mean orientation of
       platelets. Hence, $\theta_{1}=\theta_{2}=1/2$
      corresponds to a random platelet orientation in the $x_{1}x_{2}$-plane.}
\label{fig:Hydride_orientation}
\end{figure}

We consider a material with precipitate volume fraction $\zeta$ and
precipitate mean orientation characterized by the parameters
$\theta_{i}$, which are the fractions of platelets aligned with their
principal axis along the coordinate directions $x_i$; see Fig.
\ref{fig:Hydride_orientation} for a two-dimensional
representation. The volume fraction of platelets aligned with the i:th
coordinate axis, $\zeta_{i}$, is then simply $\zeta_{i}=\zeta
\theta_{i}$. Hence, its time derivative is
\begin{equation}
  \label{eq:Kappa_i1}
  \frac{d\zeta_{i}}{dt} = \zeta \frac{d \theta_{i}}{dt} +
  \theta_{i} \frac{d \zeta}{dt}.
\end{equation}
Next, we make the ansatz
\begin{equation}
  \label{eq:Kappa_i2}
  \frac{d\zeta_{i}}{dt} = n_{i} \frac{d \zeta}{dt},  \; \;
  \text{if} \; \; \frac{d\zeta}{dt} > 0,
\end{equation}
where $n_{i}$ is the fraction of platelets nucleated in the
$i$-direction at a given instant. $n_{i}$ is a function of the current
temperature and stress state, and it also depends on the density of
precipitate habit planes aligned in the $i$:th direction. The latter
quantity is linked to microstructural properties of the material,
e.g. the texture and grain shape. Combining Eqs. (\ref{eq:Kappa_i1})
and (\ref{eq:Kappa_i2}) gives
\begin{equation}
  \label{eq:Theta_i}
  \frac{d\theta_{i}}{dt} = \frac{n_{i}-\theta_{i}}{\zeta} \frac{d
  \zeta}{dt}, \; \; \text{if} \; \; \frac{d\zeta}{dt} > 0.
\end{equation}
Equation (\ref{eq:Theta_i}) provides the evolution laws for the
precipitate orientation parameters $\theta_{i}$.  It should be emphasized
that $d\theta_{i}/dt \neq 0$ only when $d\zeta/dt > 0$, i.e. the
precipitate mean orientation is assumed to change as a result of nucleation
only. If precipitates dissolve, then $d\zeta/dt < 0$, and $\theta_{i}$ are
unaffected. This means that the rate of dissolution is supposed to be
the same for all precipitates, irrespective of their orientation.

 From classical nucleation theory, the fraction of platelets nucleated
in the $i$-direction is calculated as (\cite{Puls_1985} and in
particular appendix to \cite{Jernkvist_Massih_2008})
\begin{equation}
\label{eq:n_j}
n_{i}(T,\sigma_{kl}) =  \Big(1 + m_{i0}\;e^{-\beta\Omega^\ast\sigma_{kl}\Delta \epsilon_{kl}^{T}}\Big)^{-1},
\end{equation}
where $m_{i0}=(1-n_{i0})/n_{i0}$; $n_{i0}$ is the fraction of
platelets nucleated in the $i$-direction under stress-free conditions,
$\Omega^\ast$ the critical volume for nucleation of a new phase,
$\beta=(k_BT)^{-1}$, $T$ the absolute temperature, $k_B$ the Boltzmann
constant, $\sigma_{kl}$ the stress tensor, and $\Delta
\epsilon_{kl}^{T}$ the differential misfit (transformation) strain
tensor. For an oblate spheroid with major and minor radii $a$ and $c$,
$\Omega^\ast=4\pi a^{\ast 2}c{^\ast}/3$, where the asterisk symbol signifies the
critical values for nucleation.

Under constant temperature and stress, Eq.  (\ref{eq:Theta_i}) can be solved,
subject to an initial condition $\theta_i(t=0)=\theta_{i0}$, yielding
\begin{equation}
  \label{eq:Theta_i-sol}
  \theta_{i}(t) =
n_i-\frac{n_{i}-\theta_{ie}}{1-(1-\zeta_0/\zeta_e)e^{-t/\tau}},
\end{equation}
\noindent
where we utilized Eq. (\ref{eq:Hydride_fraction_sol}) with $\tau\equiv \tau_0$
and $\theta_{ie}$ is $\theta_{i}$ at
equilibrium, viz.
\begin{equation}
  \label{eq:Theta_i-eq}
  \theta_{ie}= n_i-(n_i-\theta_{i0})\frac{\zeta_0}{\zeta_e}.
\end{equation}
\noindent
In a plane, $\theta_{2}=1-\theta_{1}$, and  therefore a single
differential equation is  sufficient to describe the change
in the precipitate mean orientation.

Let us now consider a tubular geometry in which the precipitate orientation is
referred to the polar coordinates $(r,\varphi)$, with
$\theta=(\theta_r,\theta_\varphi)$ and $n=(n_r,n_\varphi)$. The
differential misfit (transformation)  strain tensor is then defined
through
\begin{equation}
  \label{eq:Diff_mf_strain_tensor}
  \Delta \epsilon_{kl}^{T} = \epsilon_{kl}^{Tr} - \epsilon_{kl}^{T\varphi},
\end{equation}
where the right-hand-side terms are the unconstrained transformation
strain tensors for radially and circumferentially aligned
precipitates, respectively.  We note that the fraction of precipitates
nucleated in the radial direction under stress-free conditions,
$n_{r0}$, is related to the fraction of precipitate habit planes
parallel to the radial direction. This depends on microstructural
properties of the material, such as crystallographic texture and grain
shape. For Eq. (\ref{eq:n_j}) to be non-trivial, we require that
$0<n_{r0}<1$.

\begin{figure}[htbp]
  \begin{center}
    \includegraphics[width=0.70\textwidth]{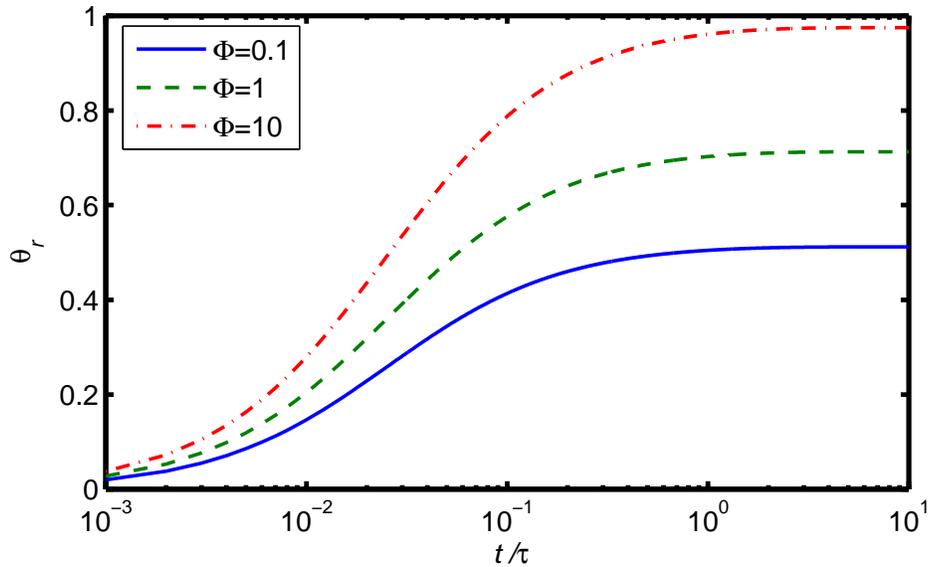}{} \\
  \end{center}
 \caption{Generic (scaled) plots of platelet orientation vs. time at
different energy scales $\Phi$, defined in the text, using Eqs. (\ref{eq:Theta_i-sol})-(\ref{eq:n_j_fd}).}
 \label{fig:theta-t}
\end{figure}

Let us write  Eq. (\ref{eq:n_j}) in a compact form
\begin{equation}
\label{eq:n_j_fd}
n_{i} =  \big(1 + e^{-\Phi_i}\big)^{-1},
\end{equation}
where $\Phi_i \equiv A_{n}\sigma_{kl}\Delta\epsilon_{kl}^{T}/T-\ln{(m_{i0})}$ is a scaled
energy parameter. In Fig. \ref{fig:theta-t}, we have plotted $\theta_r$ as
a function of the reduced time ($t/\tau$) and  parameter
$\Phi_i$ using Eqs. (\ref{eq:Theta_i-sol})-(\ref{eq:n_j_fd}). In this
calculation, we have assumed $\zeta_0/\zeta_e=0.025$ and
$\theta_{i0}=0$.

\section{Precipitation of hydrides in zirconium alloys}
\label{sec:zr-alloy}

Past theoretical studies on the effect of hydride reorientation in
zirconium alloys have shown that the preferential orientation of
hydride precipitates under stress is most effective during the
nucleation stage of the precipitation \cite{Ells_1970,Puls_1985}. It
is argued that the driving force for orienting under stress can be a
substantial fraction of the overall force during
nucleation. Furthermore, it is  indicated that hydride growth is
unlikely to play a role in preferential orientation
\cite{Ells_1970,Puls_1985}.

\begin{table}
  \caption{Model parameters for Zr2.5Nb alloy, where $T$ is the absolute temperature.}
\label{tab:param}
\begin{center}
  \begin{tabular}{lll}
  \hline
  Parameter  &  unit & source  \\ \hline
 $C_L^d = 8.080\times 10^4\exp[-4151.8/T]$ & wppm &
\cite{Pan_et_al_1996}  \\
 $C_L^p = 2.473\times 10^4\exp[-3107.9/T]$ & wppm & \cite{Pan_et_al_1996}  \\
 $C_U = 16641-37465 \exp[-2414/T]$ & wppm & \cite{Zuzek_et_al_2000}
\\
$D_\alpha = 1.17\times 10^{-7}\exp[-4041.2/T]$ & m$^2$/s & \cite{Sawatzky_et_al_1981}  \\

\hline
   \end{tabular}
\end{center}
\end{table}

Let us first identify material-dependent parameters in the model described
in the foregoing section for the case of hydrogen in a Zr-alloy,
namely Zr-2.5wt\% Nb (Zr-2.5Nb), \cite{Northwood_&_Kosasih_1983}. The
phases in the binary Zr-H system are as following
\cite{Zuzek_et_al_2000}: There are two allotropic forms of zirconium,
namely hexagonal closed-pack (hcp) $\alpha$-Zr and body-centered cubic
(bcc) $\beta$-Zr, two stable hydride phases, face-centered cubic (fcc)
$\delta$-hydride ($\approx$ ZrH$_{1.5}$) and face-centered tetragonal
(fct) $\varepsilon$-hydride ($\approx$ ZrH$_2$). In addition, there is
one metastable hydride phase, $\gamma$-hydride, with fct structure ($\approx$
ZrH), which can exist at the lower temperatures in the
($\alpha+\delta$)-phase region. The $\delta$-hydrides appear usually
as plate-like (oblate spheroid) particles, whereas the
$\gamma$-hydrides are commonly observed as needle-like ellipsoidal
particles.

We first regard the equilibrium hydride volume fraction $\zeta_e$ as expressed
by Eq. (\ref{eq:Lever_rule}). Two material-dependent parameters appear
in this equation, i.e. $C_L$ and $C_U$. They  are considered to be functions of
temperature   only (table \ref{tab:param}), despite a slight pressure
dependence of these quantities. We should also note that $C_L$ differs for hydride
dissolution  $C_L^d$ and hydride precipitation $C_L^p$
\cite{Pan_et_al_1996}. The $\delta$-line phase boundary $C_U$ was
obtained by an exponential fitting to the data for the H-Zr system presented in
\cite{Zuzek_et_al_2000}.  We  next estimate the relaxation time $\tau$ in
Eq. (\ref{eq:Theta_i-sol}). As noted earlier, $\tau \propto
\ell^{2}/D_{\alpha}$ where $\ell$  is the observed distance between hydrides and
$D_{\alpha}$ the hydrogen diffusivity in $\alpha$-phase
zirconium (table \ref{tab:param}). For small hydride volume fractions, $\ell$ can
be related to platelet thickness $h$ via \cite{Kearns_1968}
\begin{equation}
  \label{eq:hydride-dist}
  \ell = \frac{h}{2}\Big(\frac{V_m}{\zeta}\Big) \approx \frac{1}{2}\frac{h}{\zeta}.
\end{equation}
\noindent
where $V_m$ and $\zeta$ are the volume fractions of the metal and
hydride, respectively. Moreover, in a Zr-H alloy, Kearns
\cite{Kearns_1958} observed that for hydrogen concentrations up to
1000 weight parts per million (wppm), $\ell = 6.7\times 10^3 h/C$,
where $C$ is wppm hydrogen. With this relation, the relaxation time is
determined from measured values of $h$ rather than $\ell$. For
example, for $C=100$ wppm, $h=0.5$ $\mu$m (typical observed value),
$T=580$ K and $D_\alpha$ given in table \ref{tab:param}, we calculate
$\tau \approx 10$ s. Kearns \cite{Kearns_1958} found that a
dissolution kinetics model similar to that in Eq.
(\ref{eq:Hydride_fraction_app}) fitted well his experimental data. He
measured the time needed for complete dissolution of hydrides at
constant temperatures, ranging from 570 to 690 K. The hydrogen content
of his Zircaloy-4 samples was from 60 to 205 wppm.

\begin{figure}[htbp]
  \begin{center}
    \includegraphics[width=0.65\textwidth]{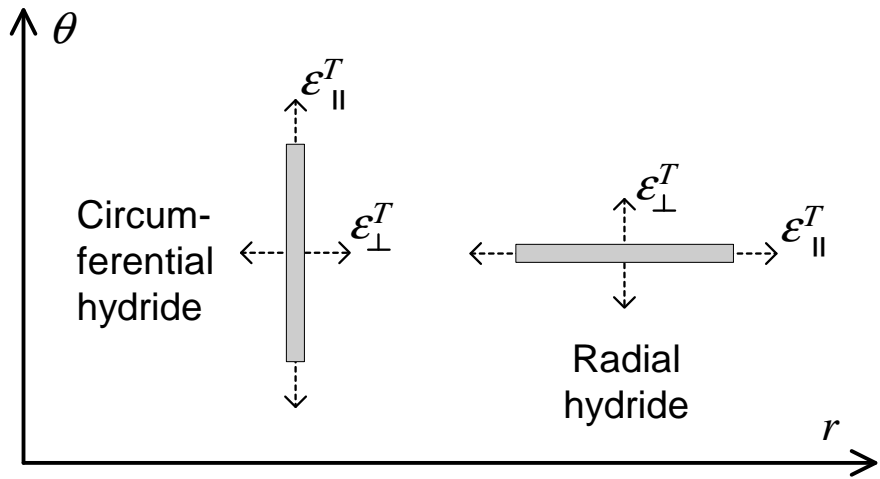}{} \\
  \end{center}
 \caption{Radial and circumferential hydrides. The hydride
transformation (misfit) strain is $\epsilon_{\perp}^{T}$ perpendicular to the hydride
platelets, and $\epsilon_{\parallel}^{T}$ parallel to the platelets.}
   \label{fig:Hydride_rt}
\end{figure}

To calculate the fraction of hydrides nucleated in a given direction
using Eq. (\ref{eq:n_j}), we need the components of the differential
misfit strain tensor in Eq.
(\ref{eq:Diff_mf_strain_tensor}). Carpenter \cite{Carpenter_1973}
calculated the unconstrained misfit strains associated with the
formation of $\delta$- and $\gamma$-hydride in $\alpha$-phase
zirconium, and expressed the strain components with respect to
crystallographic orientations of the zirconium matrix; see table
\ref{tab:Misfit_strains}. Since $\delta$-hydride platelets form on
planes, which are roughly aligned with the zirconium basal plane, the
$[0001]$ direction in table \ref{tab:Misfit_strains} corresponds
approximately to the platelet thickness (normal) direction. Hence, the
misfit strain is approximately $\epsilon_{\perp}^{T}=0.072$ normal to
$\delta$-hydride platelets, and $\epsilon_{\parallel}^{T}=0.0458$
parallel to the platelets.  With these approximations, it follows from
Fig. \ref{fig:Hydride_rt} that the misfit strain components for radial
hydrides are
\begin{equation}
\label{eq:mfs_r}
\epsilon_{rr}^{Tr} = \epsilon_{\parallel}^{T}, \; \;
\epsilon_{\varphi\varphi}^{Tr} = \epsilon_{\perp}^{T}, \; \;
\epsilon_{zz}^{Tr} = \epsilon_{\parallel}^{T}.
\end{equation}
Likewise, for circumferential hydrides
\begin{equation}
\label{eq:mfs_t}
\epsilon_{rr}^{T\varphi} = \epsilon_{\perp}^{T}, \; \;
\epsilon_{\varphi\varphi}^{T\varphi} = \epsilon_{\parallel}^{T}, \; \;
\epsilon_{zz}^{T\varphi} = \epsilon_{\parallel}^{T}.
\end{equation}
Accordingly, the components of the differential misfit strain tensor
in Eq. (\ref{eq:Diff_mf_strain_tensor}) become: $\Delta
\epsilon_{rr}^{T} = \epsilon_{\parallel}^{T} - \epsilon_{\perp}^{T} =
- 0.0262$, $\Delta \epsilon_{\varphi\varphi}^{T} =
\epsilon_{\perp}^{T} - \epsilon_{\parallel}^{T} = 0.0262 $, and
$\Delta \epsilon_{zz}^{T} = \epsilon_{\parallel}^{T} -
\epsilon_{\parallel}^{T} = 0. $ Introducing $\Delta \epsilon_{o}^{T} =
\epsilon_{\perp}^{T} - \epsilon_{\parallel}^{T} = 0.0262$ and these
relations into Eq. (\ref{eq:n_j_fd}), we find $n_{r} =
(1+\exp[-\Phi_r])^{-1}$, with $\Phi_r=\Omega^\ast\Delta
\epsilon_{0}^{T}(\sigma_{\varphi\varphi} - \sigma_{rr})/k_BT
-\ln(m_{r0})$. This relation for $n_r$ can be used for cylindrical
geometries to calculate the fraction of $\delta$-hydride platelets
that nucleate with a radial orientation, given the stress state and
temperature in the material. We may treat $m_{r0}$ and $\Omega^\ast$
as material dependent input parameters to the model, which can be
determined from hydride reorientation experiments on a particular
material. For example, Hardie and Shanahan's (HS) stress reorientation
experiment on Zr-2.5Nb containing 100 wppm hydrogen (cf.  Fig. 8 of
\cite{Hardie_Shanahan_1975}, samples cooled from 400$^\circ$C) gives
$\Omega^\ast/k_B = 1.212 \times 10^{-3}$ KPa$^{-1}$ and
$m_{r0}=1041$. Employing these constants for
$\sigma_{\varphi\varphi}=150$ MPa, $\sigma_{rr}=0$, $T=573$ K, we
obtain $\Phi\approx 1.4$ (cf. Fig. \ref{fig:theta-t}). We should note
that Eq. (\ref{eq:n_j}) is formulated in terms of true stress, which
means that the stress tensor $\sigma_{kl}$ comprises contributions
from applied loads as well as possible residual stresses, thermally
induced stresses, etc.  Hydride reorientation experiments show that
residual stresses can have a significant impact on hydride orientation
\cite{Leger_Donner_1985,Singh_et_al_2006}.

\begin{table}[!htb]
  \begin{center}
    \caption{Zirconium hydride misfit strains, as reported by Carpenter \cite{Carpenter_1973}.}
    \begin{tabular}[c]{c|cc}
      \hline
      Direction in $\alpha$-phase & \multicolumn{2}{c}{Unconstrained
      misfit strain } \\
      zirconium matrix & $\delta$-hydride &  $\gamma$-hydride \\
      \hline
      $[0001]$         & 0.0720 & 0.0570 \\
      $[11\bar{2}0]$   & 0.0458 & 0.0055 \\
      $[1\bar{1}00]$   & 0.0458 & 0.0564 \\
      \hline
    \end{tabular}
    \label{tab:Misfit_strains}
  \end{center}
\end{table}
%

We now rewrite explicitly the expression for the number of platelets nucleated in
the radial direction, Eq. (\ref{eq:n_j}), for $\sigma_{rr}=0$, namely
\begin{equation}
n_r = \Big(1+m_{r0}\exp[-\beta\Omega^\ast\Delta\epsilon_{0}^{T}\sigma_{\varphi\varphi}]\Big)^{-1}.
\label{eqn:n_r}
 \end{equation}
\noindent
It is important to point out that the supposition that the parameter
$\Omega^\ast$ is a material constant at best is a rough
approximation. There exists a supercooling effect, which shows that
the formation of a new phase occurs at a temperature below the
equilibrium solvus temperature $T_e$ by a deviation $\Delta T =
T_e-T$. In other words, the critical volume $\Omega^\ast$ or the Gibbs
energy of formation $G^\ast$ are decreasing functions of $\Delta T =
T_e-T$ (see e.g. \cite{Porter_Easterling_1981}). Furthermore, it can
be shown that for a plate-like oblate spheroid precipitate (see
appendix to \cite{Jernkvist_Massih_2008})
\begin{equation}
\Omega^*
=9\pi\gamma_f\gamma_e^2g_n^{-3},
\label{eqn:critical-vol-app}
 \end{equation}
\noindent
and the critical energy for nucleation
\begin{equation}
G^* =\frac{9\pi}{2}\,\gamma_f\gamma_e^2\; g_n^{-2},
\label{eqn:critical-energy}
 \end{equation}
\noindent
with the nucleation energy density expressed as
\begin{equation}
g_n =k_BTC_x\ln(C_s/C_{eq})-g_{\epsilon}+g_{ex}.
\label{eqn:tot-energy-density}
 \end{equation}
\noindent
Here $C_x$, $C_s$, $C_{eq}$ are the concentrations of the solute
(hydrogen) in the precipitate (hydride), in the supersaturated matrix
(Zr), and in the matrix in equilibrium with the precipitate,
respectively. Also, $g_{\epsilon}$ is the misfit strain energy
(self-energy) per unit volume, and
$\gamma_f$, $\gamma_e$ are the specific interface energies of the flat
and the edge side of the platelet, respectively
\cite{Sauthoff_1976}.
The interfacial energy may be estimated from  measured data on
nucleation rate \cite{Sauthoff_1981}, but to our knowledge such data are unavailable for
the hydride-Zr system. Furthermore, we should note that hydrogen in the matrix
depletes during the formation of a new phase and therefore $C_s$ is a
time-dependent variable. More precisely, we may relate $C_s$ to the
hydride volume fraction and the total hydrogen concentration in the
material $C$ in the manner
\begin{equation}
C = C_s(1-\zeta)+\zeta C_U,
\label{eqn:mix-conc}
 \end{equation}
\noindent
where $\zeta$ is governed by
Eq. (\ref{eq:Hydride_fraction_app}). Finally, for the calculation of
the relaxation time $\tau$, we relate the inter-hydride distance in
the specimen to the volume fraction of hydride by assuming
\begin{equation}
\ell = \frac{\ell_0\ell_1}{\ell_1+\zeta(\ell_0-\ell_1)},
\label{eqn:mix-el}
 \end{equation}
\noindent
where $\ell_0$ and $\ell_1$ are observational constants. Thus to
account for the effect of supercooling, the two ordinary differential
equations (\ref{eq:Hydride_fraction_app}) and (\ref{eq:Theta_i}) need
to be solved simultaneously, by making use of
Eqs. (\ref{eq:Lever_rule}) and (\ref{eqn:n_r})-(\ref{eqn:mix-el}).

Using the aforementioned method, we attempt to simulate an experiment
by Hardie and Shanahan \cite{Hardie_Shanahan_1975} on stress
orientation of hydrides in Zr-2.5Nb specimens under non-isothermal
conditions. More specifically, we consider a case in HS's series C
tests where the specimen contained $C=100$ wppm hydrogen. The specimen
was heated from room temperature to 673 K to dissolve the hydrides,
then cooled under a constant stress (150 MPa) at rates 3.3 Ks$^{-1}$
(673 to 573 K), 2.5 Ks$^{-1}$ (573 to 473 K) and 1.7 Ks$^{-1}$ (473 to
373 K). In our computations, the variable $C_L$ in
Eq. (\ref{eq:Lever_rule}) is the hydrogen solubility for precipitation
$C_p$. Our analyses show that the choice of $C_L$ has a strong impact
on the calculated results, since it determines the onset of hydride
precipitation during a cooling sequence. Therefore, we assume that
$C_L=S_fC_p$, where $S_f$ is a scaling factor, $1<S_f<2$, and $C_p$ is
given by the expression for $C_L^p$ in table \ref{tab:param}. We have
used the following set of values for the model parameters:
$g_{ex}=0.0262\sigma_{\varphi\varphi}$, $g_\epsilon=100$ MJm$^{-3}$,
$\gamma_f=0.065$ Jm$^{-2}$, $\gamma_e=0.28$ Jm$^{-2}$, $l_0=100$
$\mu$m, $l_1=0.1$ $\mu$m, $S_f=1.47$, and $n_{r0} \approx 5\times
10^{-4}$. In addition, we consider $\delta$-hydride with
$C_x=xN_A/\bar{V}_h$, $x=1.66$, the molar volume
$\bar{V}_h=1.63\times10^{-5}$ m$^3$mol$^{-1}$, and $N_A$ the Avogadro
constant. Moreover, we have assumed $C_{eq}=C_{L}^d(T)$ as given in
table \ref{tab:param}.

\begin{figure}[htbp]
 \begin{center}
\begin{tabular}{cl}
\includegraphics[width=0.55\textwidth]{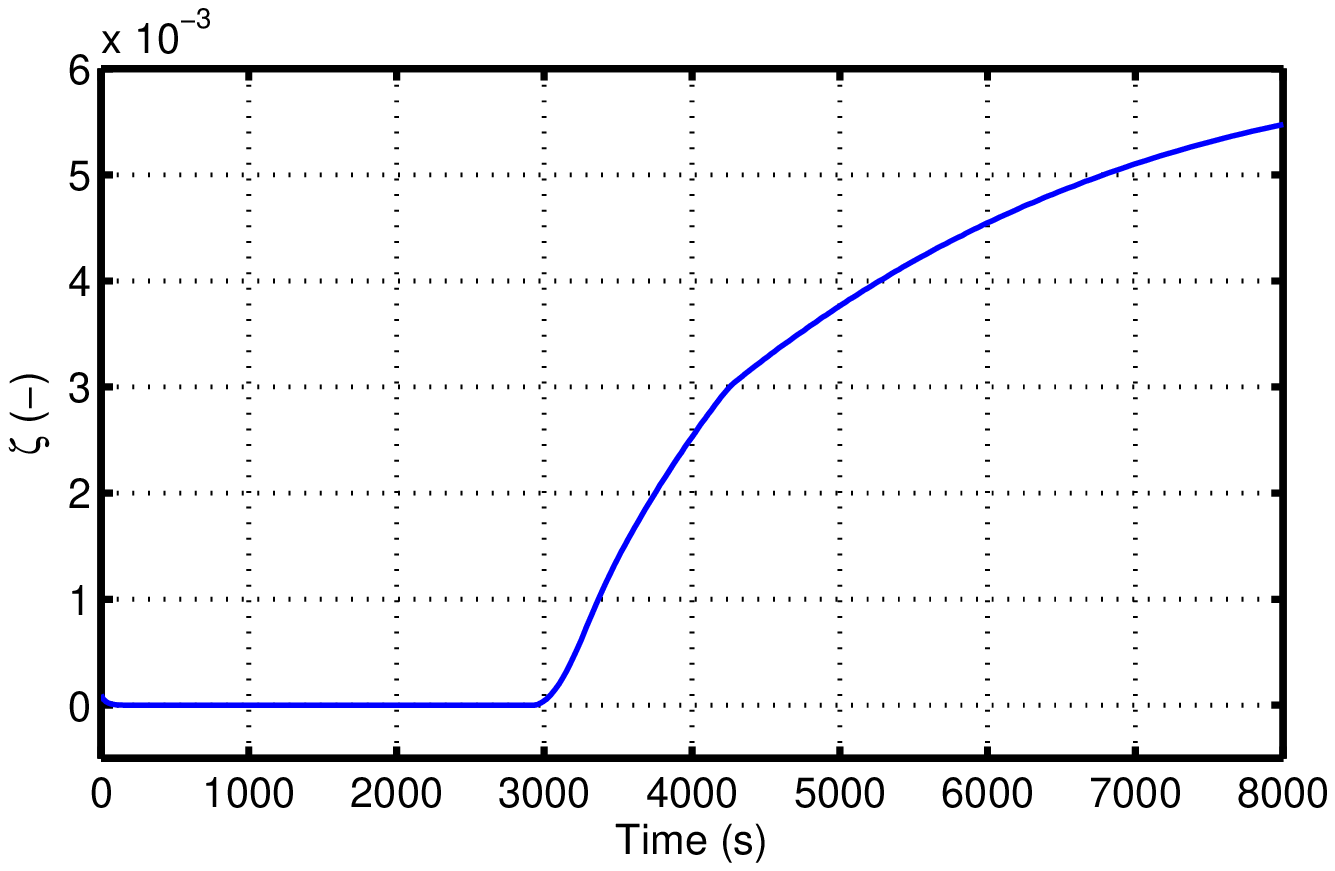}{} \\
\includegraphics[width=0.75\textwidth]{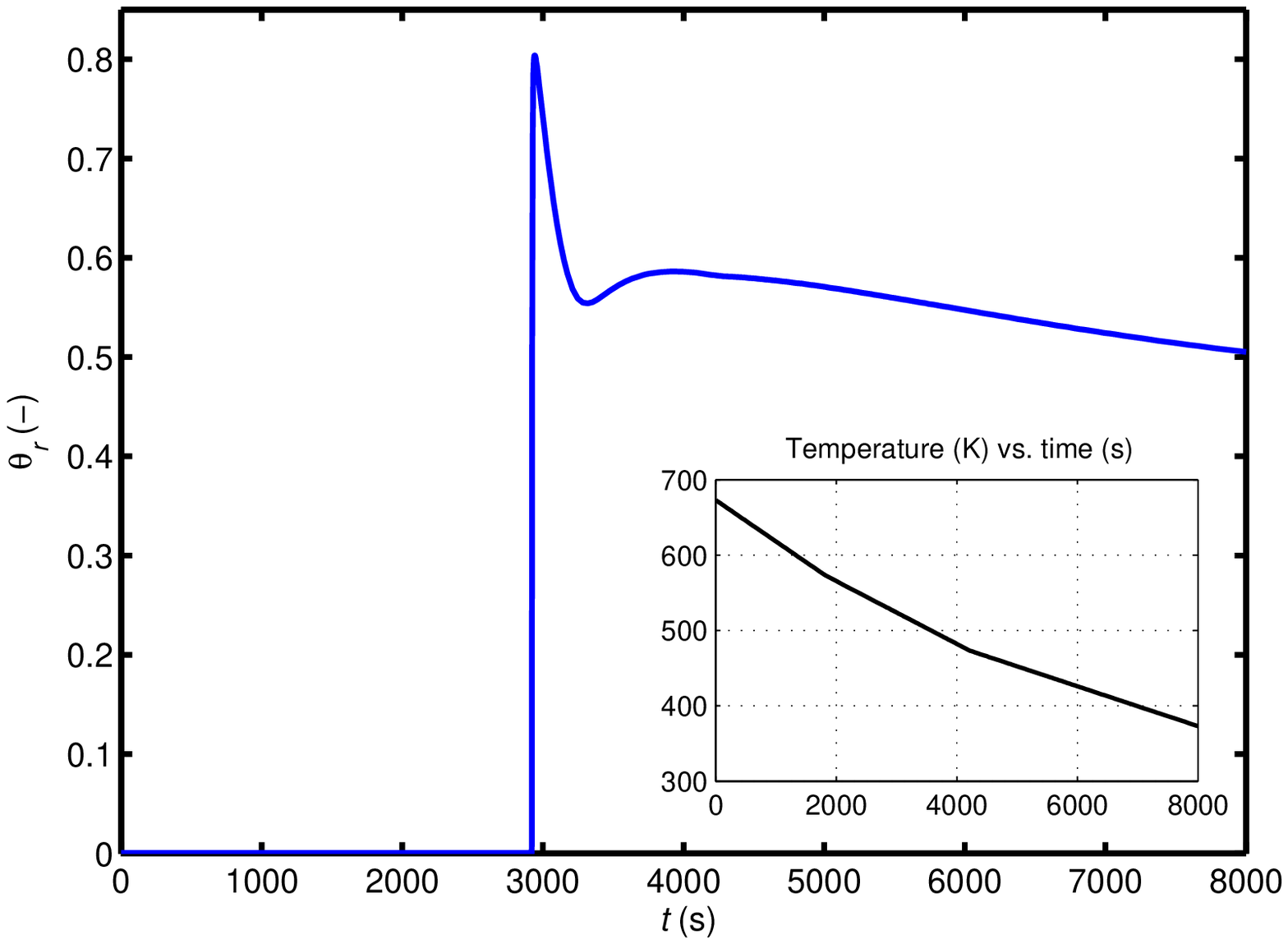}{} \\
\end{tabular}
  \end{center}
 \caption{Results of calculations of the hydride volume fraction ($\zeta$) and the
orientation parameter ($\theta_r$) as a function of
time at a constant applied stress of 150 MPa in  Zr-2.5Nb alloy. The sample contained 100 wppm
hydrogen. The inset figure displays the cooling history during reorientation.}
   \label{fig:theta-hist}
\end{figure}

We have solved the system of the aforementioned equations by using the
Runge-Kutta algorithm of order 4 and 5
\cite{Quarteroni_Saleri_2003}. The results for the time variation of
the hydride volume fraction $\zeta$ and the orientation parameter
$\theta_r$ are depicted in Fig.  \ref{fig:theta-hist}.  It is seen
that nucleation of hydride occurs immediately around 3000 s after the
start of cooling, corresponding to the temperature of about 525
K. Then $\theta_r$, after a sharp dip to a shallow minimum, raises and
falls slowly to a near equilibrium value of $\theta_r=0.5$, at
$t=8000$ s, $T=373$ K. This value is close to the experimental result
of Hardie and Shanahan (cf. Fig. 8 of \cite{Hardie_Shanahan_1975}). We
have also tabulated the results of computations at $T=373$ K for
several initial (total) hydrogen concentrations: $C=[100,200,300]$
wppm, and applied stresses: $\sigma_{\varphi\varphi}=100-300$
MPa. Table \ref{tab:theta-eq} shows these results for $\theta_r$. The
second column in this table can be compared with HS's experimental
data (Fig. 8 of \cite{Hardie_Shanahan_1975}), which shows a reasonable
agreement. The corresponding calculated hydride volume fractions for
the three hydrogen concentrations are:
$\zeta=[0.0055,0.0115,0.0176]$. Since the nucleation energy density
$g_n$ is the driving force for hydride formation, we have listed their
calculated values in table \ref{tab:trans-energy}. For example, at
$C=100$ wppm, $\sigma_{\varphi\varphi}=150$ MPa, $g_n= 554$
MJm$^{-3}$, and so on.

\begin{table}[!htb]
  \begin{center}
    \caption{Calculated orientation parameter $\theta_r$ for hydride
formation at  various total hydrogen contents $C$ and applied stresses
$\sigma_{\varphi\varphi}$  at $T=373$ K.}
    \begin{tabular}[c]{cccc}\hline
  $\theta_r$ (-)     & & $C$ (wppm) &\\\hline
   $\sigma_{\varphi\varphi}$ (MPa) & 100 & 200 &  300 \\\hline
  100 & 0.0765 & 0.2381 & 0.4067
\\
  150 & 0.5052 & 0.8071 & 0.8461
\\
  200 & 0.9153 & 0.9772 & 0.9797
\\
  250 & 0.9912 & 0.9980 & 0.9982
\\
  300 & 0.9991 & 0.9998 & 0.9998
\\
\hline
 \end{tabular}
 \label{tab:theta-eq}
\end{center}
\end{table}

\begin{table}[!htb]
  \begin{center}
    \caption{Calculated nucleation energy densities $g_n$ for hydride
formation at various total hydrogen contents $C$ and applied stresses
$\sigma_{\varphi\varphi}$  at $T=373$ K.}
    \begin{tabular}[c]{cccc}\hline
  $g_n$ (MJm$^{-3}$)     & & $C$ (wppm) &\\\hline
   $\sigma_{\varphi\varphi}$ (MPa) & 100 & 200 &  300 \\\hline
 100 & 552 & 540 & 537\\
  150 & 554 & 541 & 540\\
  200 & 555 & 543 & 540\\
  250 & 556 & 543 & 541\\
  300 & 558 & 545 & 542\\
 \hline
 \end{tabular}
 \label{tab:trans-energy}
\end{center}
\end{table}
\section{Discussion}
\label{sec:discuss}

 The behaviour seen in Fig.  \ref{fig:theta-hist} may be interpreted
as an expression of supercooling effect, whereupon a metastable
configuration (local minimum) appears before the system settles down
to a more stable configuration. This effect, despite its importance,
to our knowledge, has not been studied experimentally in a
quantitative fashion for the Zr-H system. In our study as in the
earlier investigation \cite{Puls_1985}, we assumed that stress
orientation of hydrides occur primarily during the nucleation stage of
precipitation. The kinetics of nucleation is characterized by the
nucleation rate. The steady-state nucleation rate $J$ gives the number
of (hydride) nuclei forming per unit volume and time, expressed as
\cite{BAC_2005}, $J=Z\nu_cN\exp(-G^*/k_BT)$, where $Z$ is the
Zeldovich factor ($\approx 0.1$), $\nu_c$ the frequency factor,
i.e. the rate atoms add to the critical nucleus, $N$ the number of
sites available for nucleation, and $G^\ast$ is given by
Eq. (\ref{eqn:critical-energy}).  Note that $G^\ast \propto 1/g_n^2$,
so a slight increase in $g_n$ gives a significant rise to the
nucleation rate.

Nucleation experiments similar to that of Tanaka et
al. \cite{Tanaka_et_al_1978} made on the Fe-N system, are valuable to
determine the hydride nucleation rate as well as the degree of
hydride orienting as a function of hydrogen concentration and applied
stress. Such experiments would allow to identify or verify some of the model
parameters used in the computations here.

Finally, an issue worth addressing is the role of late-stage growth or
coarsening on hydride stress orienting. The late-stage growth follows
the Lifshitz-Slyozov law \cite{Lifshitz_Slyozov_1961}, expressed as
\begin{equation}
\bar{a}^3- \bar{a}_0^3 = \alpha_D(t-t_0),
\label{eqn:LSW}
 \end{equation}
\noindent
where $\bar{a}$ is the mean particle (precipitate) radius at time $t$,
$\bar{a}_0$ the mean particle radius at time $t_0$ when coarsening
commences, and $\alpha_D$ is the rate constant for diffusion-limited
coarsening. Such a behaviour has been observed for growth of hydride
platelets in Zr-2.5Nb alloy \cite{Shalabi_Meneley_1990}. It can be
shown that the maximum rate of particle size increase occurs at
$a_{max}=2\bar{a}$, and particles with sizes $a < \bar{a}$ will
disappear with relatively high shrinkage rates.

Following the arguments of Puls  \cite{Puls_1985}, based on a model by
Sauthoff \cite{Sauthoff_1976}, we consider a tubular specimen of Zr-2.5Nb
containing hydrides, which at time $t_0$ after nucleation half of the
hydride platelets are radially oriented and the rest are circumferentially
oriented across the tube wall. If the major radii of these platelets
are denoted by $a_r$ and $a_\theta$, respectively, Sauthoff
\cite{Sauthoff_1976} obtained a
relationship between these radii using the Gibbs-Thomson-Freundlish
relation, namely
\begin{equation}
a_r = \frac{\gamma}{e\,\Delta\epsilon_0^T\sigma_{\varphi\varphi}}(1-\varrho),
\label{eqn:sauth}
 \end{equation}
\noindent
where $e$ is the eccentricity (ratio of minor to major radii),
$\bar{\gamma}=\gamma_f+2e\gamma_e$ and $\varrho=a_r/a_\theta$. By
setting $\varrho=1/2$, Eq. (\ref{eqn:sauth}) gives a minimum mean
radius $\tilde{a}_r$ for which  even the largest circumferentially
oriented hydrides will no longer grow. Using the  data in the
foregoing section with
$\sigma_{\varphi\varphi}=150$ MPa and $e=0.23$, we obtain $\tilde{a}_r=
1.1 \times 10^{-7}$ m.

In order to attain a complete orienting in radial direction, all the
circumferential hydrides must disappear in addition to half of the
radial ones. Hence, there would be a factor of four decrease in the
total number of hydride platelets. Since during coarsening the total
precipitate volume remains practically constant, this decrease
corresponds to a factor of $(4)^{1/3} \approx 1.6$ increase in the mean
platelet radius.

We may calculate, for the case of Zr-2.5Nb discussed in the foregoing section,
the time needed to increase the minimum platelet radius by a factor of
1.6. Boyd and Nicholson \cite{Boyd_Nicholson_1971} have calculated the
coarsening rate constant $\alpha_D$ for disc-shaped precipitates,
viz.
\begin{equation}
\alpha_D = \frac{16}{9}\frac{\gamma_eD_\alpha C_{eq}\bar{V}_h}{e\pi RT}.
\label{eqn:boyd}
 \end{equation}
\noindent
Using $C_{eq}=C_{L}^d(T)$ and $D_\alpha$ in table
\ref{tab:param}, and other data listed previously,
Eq. (\ref{eqn:boyd}) yields $\alpha_D = 2.715 \times 10^{-22}$
m$^3$s$^{-1}$ at $T=523$ K. Thus to increase the minimum platelet radius of
$\tilde{a}_r= 1.1 \times 10^{-7}$ m by a factor of 1.6 would take
about 15 s at $T=523$ K, according to Eq. (\ref{eqn:LSW}). Therefore
stress orienting of hydride platelets would be possible in Zr-2.5Nb
alloy during coarsening.

\section{Conclusions}
\label{sec:conclude}
In  this paper, we have presented a generic model for
calculation of stress-induced orientation of disc-shaped precipitates in
alloys. The model is applicable to situations where the stress
orienting occurs primarily during the nucleation stage of the
precipitation process. We have identified parameters appearing in the
model for the case of hydride precipitates in a zirconium alloy. The
model can be used to compute the degree of orienting as a function of
stress, temperature and time. It also can be extended to account for
the effect of supercooling during phase transition.

\subsection*{Acknowledgments}
 The work was supported in part by the Knowledge Foundation of Sweden under the
grant number 2008/0503.

 \appendix

\section{Diffusion model for precipitation}
\label{app:A}
An idealized configuration for second-phase precipitate metal composite is
considered. The hydrides are a regular array of parallel
cylinders embedded in a metal matrix. We posit that solute
precipitation is diffusion-controlled and diffusion occurs
only in the radial direction (cylindrical symmetry). Furthermore, there is no
imposed external force on the solute atoms. The solute
concentration $C(r,t)$ is a function of space $r$ and time $t$ obeying
the diffusion equation of the form
\begin{equation}
\label{eq:diffusion_eq}
\frac{\partial C}{\partial t}=\frac{1}{r}\frac{\partial}{\partial
r}\Big(rD\frac{\partial C}{\partial r}\Big).
\end{equation}

This equation is solved subject to the following initial and boundary
conditions
\begin{align}
\label{eq:diffusion_ic}
C(r,t)  &= C_0 \qquad \text{at} \quad t=0, \\
\label{eq:diffusion_bc1}
C(r,t)  &=  C_s \qquad  \text{at} \quad r=R, \quad t\ge 0, \\
\label{eq:diffusion_bc2}
\frac{\partial C(r,t)}{\partial r} & =  0 \qquad  \;\; \text{at} \quad r=r_c, \quad t\ge 0,
\end{align}
where $R$ is the radius of the cylinder and $r_c$ that of the
cell enclosing the cylinder assumed to be impenetrable. The
impenetrability (zero flux) condition is equivalent to the requirement that
$C(r,t)$ shall have the symmetry of a regularly spaced lattice of $(\pi
r_c^2)^{-1}$ identical cylinders per unit area on which precipitation
occurs \cite{Ham_1959}.
The solution for this boundary value problem
\cite{Ham_1959,Carslaw_Jaeger_1959} is expressed as
\begin{equation}
\label{eq:diffusion_sol}
\frac{C(r,t)-C_0}{C_s-C_0}=1-\pi\sum_{n=0}^\infty\Big[1-\Big(\frac{J_0(\lambda_nR)}{J_1(\lambda_nr_c)}\Big)^2\Big]^{-1}
 e^{-D\lambda_n^2t}C_n(r),
\end{equation}
\noindent
where
\begin{equation}
\label{eq:cnr}
C_n(r) = J_0(\lambda_nr)Y_0(\lambda_nR)-Y_0(\lambda_nr)J_0(\lambda_nR),
\end{equation}
and the eigenvalues $\lambda_n$ are the roots of the transcendental equation
\begin{equation}
\label{eq:eval-eqn}
J_0(\lambda_nR)Y_1(\lambda_nr_c)-Y_0(\lambda_nR)J_1(\lambda_nr_c)=0.
\end{equation}
Here $J_n(x)$ and $Y_n(x)$ are the Bessel functions of the first and
second kind, respectively, of order $n$.

\begin{figure}[htbp]
 \begin{center}
    \includegraphics[width=0.70\textwidth]{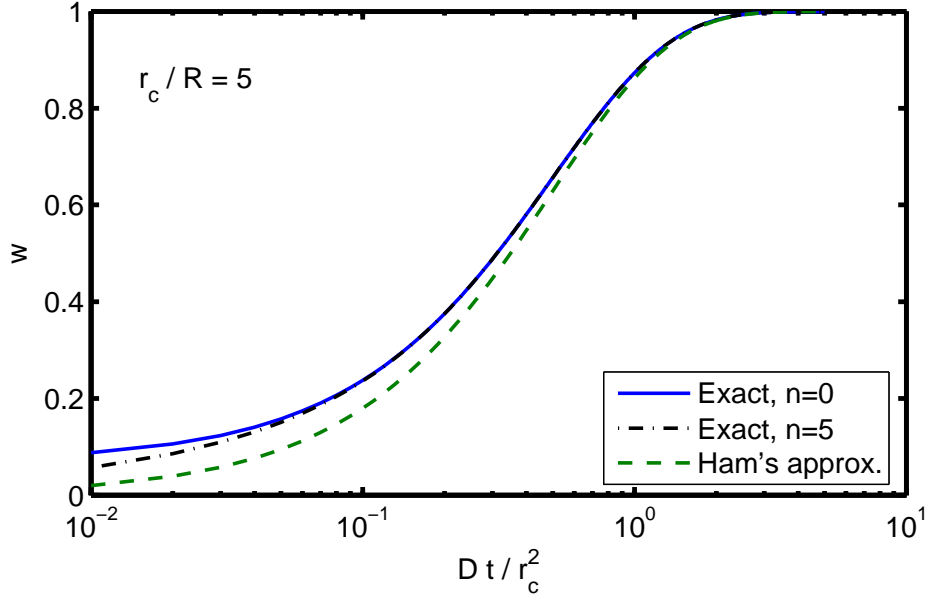}{} \\
  \end{center}
\caption{Precipitate fraction $w$ for diffusion to a cylindrical sink
of radius $R$ surrounded by an impenetrable concentric cylinder of
radius $r_c$ for $r_c/R=5$. Here ``Exact'' $n=0$ corresponds to
Eq. (\ref{eq:Ham-frac_1st-approx}), and $n=5$
to Eq. (\ref{eq:Ham-frac_exact}) with the first 6 terms in
the sum; which are compared with Ham's approximate
solution.}
\label{fig:compsol}
\end{figure}

Let us calculate the fraction of precipitates
$w$ of the excess solute as it evolves with time. This is related to
the total number of solute atoms remaining between $R$ and $r_c$,
expressed as
\begin{equation}
\label{eq:h-total}
M_t =2\pi\int_{r_c}^R r\big[C(r,t)-C_0\big]dr.
\end{equation}
Denoting the quantity of solute in the metal matrix after
infinite time by $M_\infty$, then the precipitated fraction of
particles is $w=M_t/M_\infty$. Combining  Eqs. (\ref{eq:diffusion_sol})
and (\ref{eq:h-total}), we write
\begin{equation}
\label{eq:Ham-frac_exact}
 w = 1+4\sum_{n=0}^\infty\Big[\lambda_n^2(r_c^2-R^2)\Big]^{-1}
\Big[1-\Big(\frac{J_0(\lambda_nR)}{J_1(\lambda_nr_c)}\Big)^2\Big]^{-1}
 e^{-D\lambda_n^2t},
\end{equation}
\noindent
which is essentially the result obtained by Ham \cite{Ham_1959}.
Considering now Eq. (\ref{eq:Ham-frac_exact}), we note that each
eigenmode decays exponentially with time with a different decay time
$\tau_n=r_c^2/\alpha_n^2D$. Since the large-$n$ eigenmodes decay more
rapidly with time, only the most slowly decaying eigenmode remains in
the long-time limit, namely
\begin{equation}
\label{eq:Ham-frac_1st-approx}
 w = 1-4\Big[\alpha_0^2(1-R^2/r_c^2)\Big]^{-1}
\Big[\Big(\frac{J_0(\alpha_0R/r_c)}{J_1(\alpha_0)}\Big)^2-1\Big]^{-1} e^{-t/\tau_0}.
\end{equation}
\noindent
This relation can be compared with Ham's approximate expression for $w$,
namely $w \cong 1-\exp(-t/\tau_0)$, with $\tau_0=r_c^2/\alpha_0^2D$ and
$\alpha_0^2=2[\ln(r_c/R)-3/5]^{-1}$. Evaluations of the precipitate
fraction $w$ as a function of $Dt/r_c^2 $ made according to
Eqs. (\ref{eq:Ham-frac_1st-approx}), (\ref{eq:Ham-frac_exact}) with
six term in the sum, and Ham's approximate expression show that
Eqs. (\ref{eq:Ham-frac_1st-approx}) and (\ref{eq:Ham-frac_exact})
yield almost identical results, whereas Ham's approximate relation
gives a slightly lower $w$ for $Dt/r_c^2 \le 1$, see
Fig. \ref{fig:compsol}.

\bibliographystyle{apsrev}\bibliography{DHC}

\end{document}